\documentclass[a4paper,fleqn,usenatbib]{mnras}

\usepackage{newtxtext,newtxmath}
\usepackage[T1]{fontenc}
\usepackage{ae,aecompl}

\usepackage{hyperref}
\hypersetup{
    colorlinks,
    citecolor=blue,
    filecolor=black,
    linkcolor=blue,
    urlcolor=black
}
\usepackage{multirow}
\usepackage[dvipsnames]{xcolor} 
\usepackage{amssymb} 
\usepackage{amsmath} 
\usepackage{wasysym} 
\usepackage{subfiles} 
\usepackage{subfigure} 
\usepackage{wrapfig} 
\usepackage{caption} 
\usepackage{pdfpages} 
\usepackage{textcomp} 
\usepackage{diagbox}

\newcommand\fCK{f\textsubscript{\text{CK}}}
\newcommand\fCN{f\textsubscript{\text{CN}}}
\newcommand\Mbh{M\textsubscript{\text{bh}}}
\newcommand\fgout{f\textsubscript{\text{g,out}}}
\newcommand\Rout{R\textsubscript{\text{out}}}

\newcommand\Sigmamp{\Sigma\textsubscript{\text{mp}}}

\newcommand\Teff{T\textsubscript{\text{eff}}}

\newcommand{\Ledd}{L_{\text{Edd}}}
\newcommand{\PhiAGN}{\Phi_{\text{AGN}}}
\newcommand{\WNH}{W_{N_H}}

\newcommand\addtag{\refstepcounter{equation}\tag{\theequation}}

\newcommand\tsup[1]{\textsuperscript{#1}}

\newcommand{\percmsqr}{\text{cm}\tsup{-2}}

\newcommand{\ergpers}{\,erg\,s$^{-1}$}

\title[Dusty Starburst Discs, CXB, \& AGN Number Counts]
  {The Shape of the Cosmic X-ray Background: Nuclear Starburst Discs and the Redshift Evolution of AGN Obscuration}
\author[R.\ Gohil and D.\ R.\ Ballantyne]
  {R.~Gohil\thanks{raj.gohil07@gmail.com} and D.~R.~Ballantyne\\Center
    for Relativistic Astrophysics, School of Physics, Georgia
    Institute of Technology, 837 State Street, Atlanta, GA 30332-0430,
    USA}

\pagerange{\pageref{firstpage}--\pageref{lastpage}}
\pubyear{2017}

\begin{document}
\label{firstpage}
\pagerange{\pageref{firstpage}--\pageref{lastpage}}
\maketitle

\begin{abstract}
A significant number of active galactic nuclei (AGNs) are observed to be hidden behind dust and gas. The distribution of material around AGNs plays an important role in modeling the cosmic X-ray background (CXB), especially the fraction of Type-2 AGNs ($f_2$). One of the possible explanations for the obscuration in Seyfert galaxies at intermediate redshift is dusty starburst discs. We compute the 2D hydrostatic structure of 768 nuclear starburst discs (NSDs) under various physical conditions and also the distribution of column density along the line of sight ($N_{\text{H}}$) associated with these discs. Then, the $N_{\text{H}}$ distribution is evolved with redshift by using the redshift dependent distribution function of input parameters. The $f_2$ shows a strong positive evolution up to $z=2$, but only a weak level of enhancement at higher $z$. The Compton-thin and Compton-thick AGN fractions associated with these starburst regions increase as $\propto (1+z)^{\delta}$ where the $\delta$ is estimated to be 1.12 and 1.45, respectively. The reflection parameter $R_f$ associated with column density $N_{\text{H}} \geq 10^{23.5}$ cm$^{-2}$ extends from 0.13 at $z=0$ to 0.58 at $z=4$. A CXB model employing this evolving $N_{\text{H}}$ distribution indicates more compact ($R_{\text{out}}<120$ pc) NSDs provide a better fit to the CXB. In addition to ``Seyfert-like'' AGNs obscured by nuclear starbursts, we predict that 40 to 60 per cent of quasars must be Compton-thick to produce the peak of the CXB spectrum within observational uncertainty. The predicted total number counts of AGNs in 8-24 keV band are in fair agreement with observations from \textit{NuSTAR}.
\end{abstract}


\section{Introduction}
The cosmic X-ray background (CXB) spans roughly 1 keV to 400 keV and the shape of its spectrum is characterized by a power law with the photon index of $\Gamma=1.4-1.52$ in the 2-10 keV band \citep{marshall80,deluca04,moretti09,cappelluti17} and a peak in the $\sim$ 20-30 keV band \citep{gruber99}. The spectrum up to a few keV is resolved mostly into point sources from the observations of \textit{XMM-Newton} and \textit{Chandra} \citep{worsley05}. Moreover, 33 to 39 per cent of the CXB in 8-24 keV band is resolved by \textit{NuSTAR} \citep{harrison16}. Many of these sources are confirmed to be active galactic nuclei (AGNs) \citep{bauer04} which suggests that the CXB may be the integrated X-ray spectrum of AGNs throughout the history of the Universe. Numerous AGN spectra exhibit the presence of dusty and gaseous regions \citep{gilli07,lawrence10}. The distribution of column density along the line of sight ($N_{\text{H}}$) is a key ingredient in modeling the CXB \citep{akylas12}, especially the fraction of Compton-thick (CK) AGNs ($\fCK$) where the reflection due to the Compton scattering becomes important. They possess a large amount of material ($N_{\text{H}}>10^{24}$ cm$^{-2}$) along the line of sight which makes them very difficult to observe even in X-ray. Moreover, a wide range of $\fCK$ from 5\% to 50\% can produce the observed CXB spectrum due to the degeneracy within modeling input parameters \citep{akylas12}. Therefore, it is essential to compute a theoretically motivated $N_{\text{H}}$ distribution.

The physical conditions, driving mechanisms, and geometrical configuration of the obscuring region in the vicinity of AGNs are very poorly known. They appear to be dependent on the AGN luminosity \citep{akylas06,hasinger08,ebrero09,burlon11,ueda14,aird15,buchner15,sazonov15} and also possibly on redshift \citep{ballantyne06,hasinger08,brightman12,iwasawa12,vito14}. Moreover, various mechanisms and physical conditions can play an important role for this observed obscuration such as a dusty atmosphere, star-formation, radiation pressure dominance, stellar wind, magnetic field, AGN feedback, and so-forth. For example, based on the study of 836 AGNs from hard X-ray Swift Burst Alert Telescope survey, \cite{ricci17} find that the radiation pressure on dust plays an important role in distributing the circumnuclear material which is mainly driven by an accretion rate. Furthermore, these regions may vary significantly from galaxy to galaxy (i.e., between Seyfert galaxies and quasars). To explain the observed properties of $N_{\text{H}}$, various modeling perspectives have been proposed: a simple uniform toroidal torus \citep{krolik86,pier92}, geometrically thick medium supported by infrared (IR) radiation pressure \citep{krolik07,dorodnitsyn11,dorodnitsyn12,chan16,dorodnitsyn16}, the turbulent pressure dominated torus \citep{wada05,watabe05}, warped/tilted discs \citep{nayakshin05,caproni06,lawrence10}, a clumpy torus \citep{honig07,nenkova08,honig10}, and nuclear starburst discs \citep{fabian98,wada02,thompson05,ballantyne08,hopkins16,gohil17}.

A great deal of work is done in modeling the observed cosmic X-ray background \citep[e.g.,][]{ueda03,treister05,ballantyne06b,gilli07,draper09,draper10}. However, the $N_{\text{H}}$ distributions used in those models are not related to any physical mechanism. Nuclear starburst discs are promising candidates to explain the obscuration in Seyfert galaxies at intermediate redshift $z\sim1$ \citep{ballantyne08,gohil17}. This is the era when there is a large gas-fraction available in galaxies \citep[i.e.,][]{narayanan12} and not long after the peak in the history of cosmic star-formation rate is observed \citep{madau14}. Moreover, one-dimensional (1D) \citep{thompson05} and two-dimensional \citep{gohil17} model of NSDs suggest that a disc can possess an inflationary atmosphere at parsec/sub-parsec scale when the grains are sublimated at mid-plane. Then, the AGN spectrum can be reprocessed by such an expanded atmosphere. Therefore, in this work, we use the 2D NSD theory of \cite{gohil17} and compute 768 NSDs across the input parameter space (disc size, Mach number, gas fraction, and black hole mass). A distribution of column density along the line of sight ($N_{\text{H}}$) is computed using these models. Afterward, the $N_{\text{H}}$ distribution is evolved by calculating redshift dependent distribution functions of the input parameters. By utilizing the evolution of the $N_{\text{H}}$ distribution, we predict the cosmic X-ray background as well as the AGN number counts in 2-8 keV and 8-24 keV bands.

Sect. \ref{sect:review} reviews the modeling aspect of the 2D NSD structure. Sect. \ref{sect:method} describes the methodology used in order to evolve the $N_{\text{H}}$ distribution. In Sect. \ref{sect:result}, we provide the results including the predicted cosmic X-ray background and also the AGN number counts in 2-8 keV and 8-24 keV bands. Sect. \ref{sect:limitation} discusses the results and compares to observations. Then, the paper is concluded in Sect. \ref{sect:conclusion}.

\begin{table}
\centering
\caption{Description of symbols used frequently in the paper.}
\label{table:symbols}
\begin{tabular}{cc}\hline\hline
Symbol & Description\\\hline\hline
$f_2$ & Fraction of Type-2 AGNs\\
$f_{2,R_f}$  & Fraction of obscured AGNs with high reflection\\
$f_{2,Q}$ & Fraction of obscured quasars\\
$\fCK$ & Fraction of Compton-thick AGNs\\
$\fCN$ & Fraction of Compton-thin AGNs\\
$\fgout$ & Gas fraction at the outer radius\\
$L$ & Bolometric luminosity\\
$\Mbh$ & A black hole mass\\
$N_{\text{H}}$ & Column density along the line of sight\\
$R_f$ & Reflection parameter\\
$\Rout$ & Size of a nuclear starburst disc\\
$P(\lambda)$ & Eddington ratio distribution\\
$\Phi_\text{AGN}$ & Distribution function of an active black hole mass\\
$\Phi_f$ & Distribution function gas fraction\\
$\theta$ & Viewing angle\\
$z$ & redshift\\
\end{tabular}
\end{table}

\section{A Brief Review on Modeling of NSDs}
\label{sect:review}
\begin{figure*}
\centerline{
\includegraphics[width=0.45\textwidth]{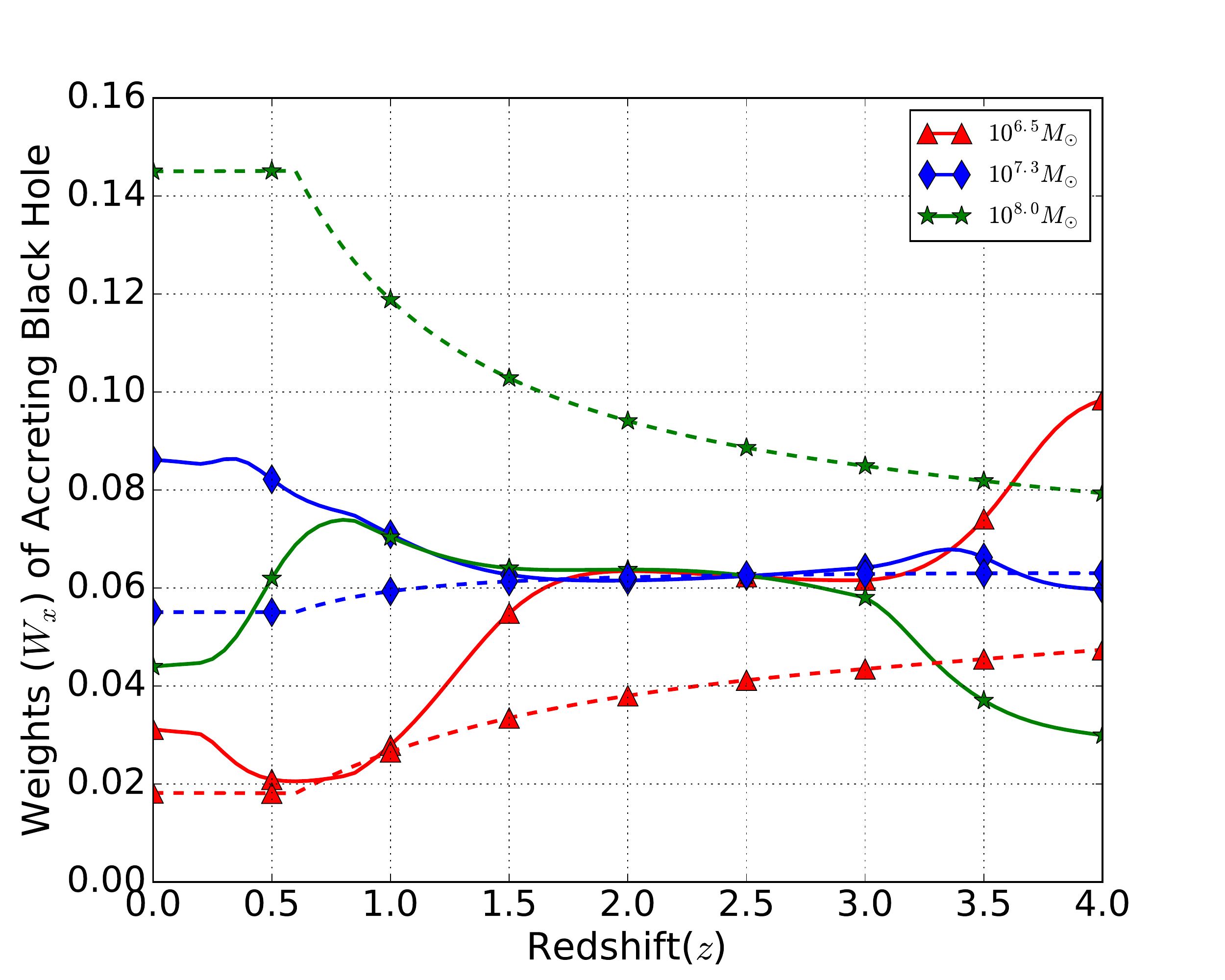}
\includegraphics[width=0.45\textwidth]{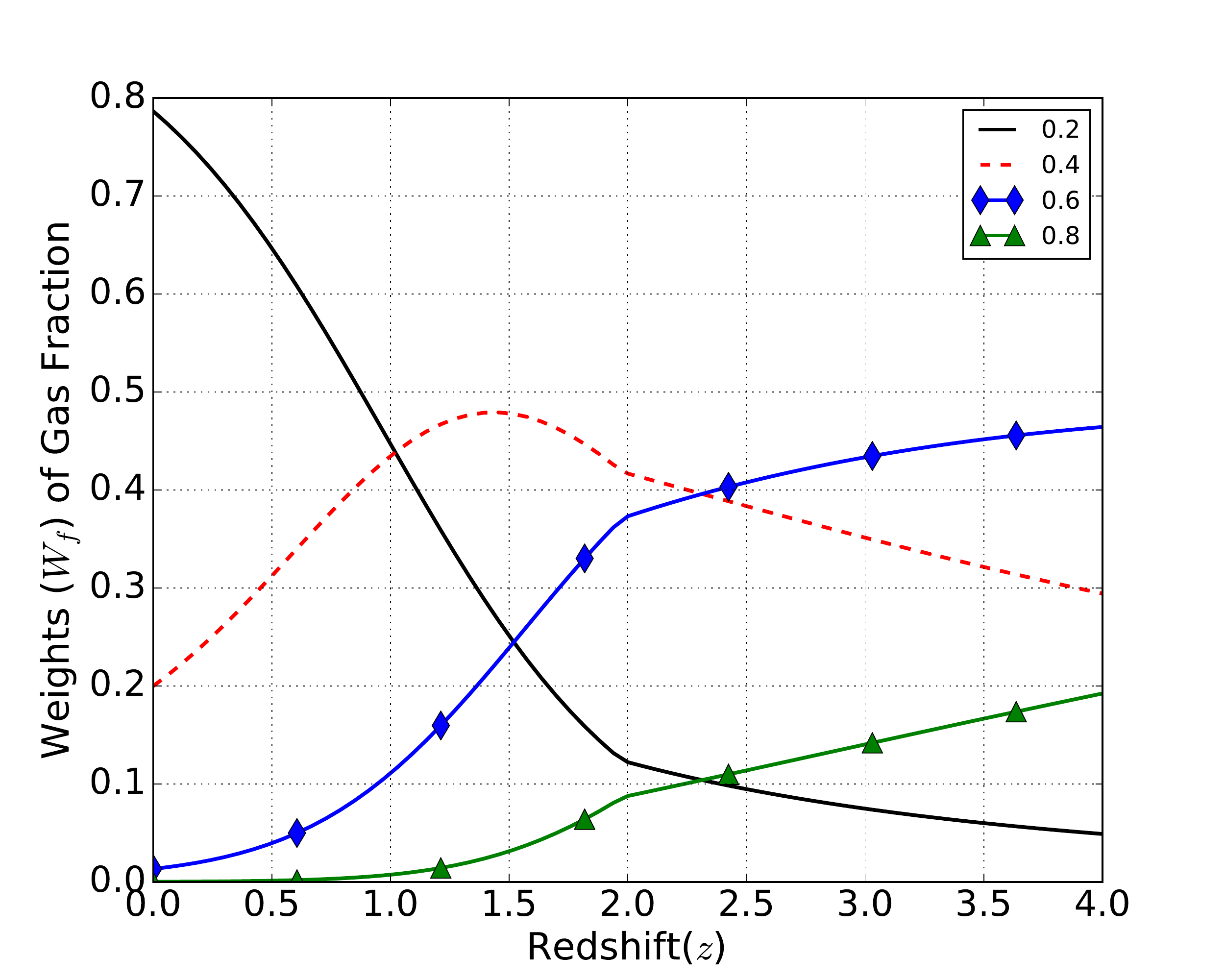}
}
\caption{\textit{Left:} Illustrating the probability weights of black hole mass $10^{6.5}M_{\astrosun}$,$10^{7.3}M_{\astrosun}$, and $10^{8.0}M_{\astrosun}$ as a function of redshift (Eq.~\ref{eqn:blackholemass}). The solid and dashed curves are computed assuming the log-normal (Eq.~\ref{eqn:lognormal}) and power-law (Eq.~\ref{eqn:powerlaw}) distribution of $P(\lambda)$, respectively. Regardless of a choice on $P(\lambda)$, weights of low-end mass accreting black holes increases with redshift and the reverse holds true for the high-end black hole mass. We have tested both the distribution functions of $P(\lambda)$ and they have a negligible effect on the evolution of $N_{\text{H}}$ distribution. \textit{Right:} The panel illustrates the probability weights $W_f$ of 20\%, 40\%, 60\%, and 80\% gas fraction as a function of redshift (Eq.~\ref{eqn:gasfraction}) which are represented by black, red, blue, and green curves, respectively. In general, the figure follows the observational feature that the gas fraction increases with redshift.}
\label{fig:wx_z}
\end{figure*}

\cite{gohil17} predicted the integrated $N_{\text{H}}$ distribution associated with the starburst regions by modeling the 2D hydrostatic structure of NSDs. The modeling of a NSD depends on four input parameters: disc size $\Rout$, the black hole mass $\Mbh$, the gas fraction at the outer radius $\fgout$, and the Mach number $m=v_r/c_s$, where $v_r$ and $c_s$ are the radial velocity and the speed of sound, respectively (frequently used variables in the paper are described in Table ~\ref{table:symbols}.). We assumed that the discs are symmetric around the $z=0$ plane and $z$-axis. The first step in constructing the 2D structure was to compute the radial distribution of the mass integrated column density $\Sigmamp(R)$ and the energy content which was parameterized by the effective temperature $\Teff(R)$. These were obtained by following 1D NSD model of \cite{thompson05}. The disc was then divided into a number of annuli and, for each annulus of a given $\Sigmamp$ and $\Teff$, the hydrostatic structure was obtained. The vertical structure was computed by solving coupled equations of hydrostatic balance, energy balance, and the radiative transfer using the iterative method. By resolving the structure at every annulus, we computed the 2D structure of NSDs where the physical quantities (i.e., temperature, density, and opacity) varied with the radial distance $R$ and the viewing angle $\theta$ (measured at the mid-plane).

The NSDs were modeled under various physical conditions which span a large range of the input parameter space. In total 192 models were computed and the range of their input parameters are shown in Table 2 of \cite{gohil17}. In total 99 models showed a starburst phenomenon at the parsec/sub-parsec scale (0.2-2 pc) when the discs exceeded the dust sublimation temperature at mid-plane. This causes a large opacity gradient resulting in an inflationary atmosphere. The phenomenon is more common in discs with a smaller size, a larger gas fraction, and/or high Mach number. These discs exhibit the inflated atmosphere with surface height ($h$) ranging from 0.2 to 30 $R$. Since 52 per cent of discs show a large expanded atmosphere, this suggests that the starburst phenomena could be common and these conditions can potentially obscure the incoming AGN radiation. Moreover, the $N_{\text{H}}$ distribution as a function of $\theta$ for a given NSD shows that its AGN can be observed as Type-1 (T1), Compton-thin (CN), or CK depending on the viewing angle. This conclusion is in agreement with the simple unification theory of AGNs which states the Type-1 and Type-2 AGNs are intrinsically the same objects except their orientations are different with respect to an observer \citep{antonucci93,netzer15}.

Finally, the $N_{\text{H}}$ distribution associated with a sample of 192 discs was computed based on random selection. The predicted $N_{\text{H}}$ distribution had a few features which were consistent with the observational evidence. First, the fraction of obscured AGNs ($f_2$) peaks near $10^{23}$ cm$^{-2}$ \citep{burlon11,ueda14,buchner15,sazonov15}. Second, the 2D NSD theory predicts the ratio of CN to CK AGNs to be 0.9 \citep{gohil17} which is within the required range, 0.5-1.6, of producing the 20-50 keV part of the CXB spectrum \citep{ueda14}. Third, the fraction of CK AGNs ($\fCK$) associated with the NSDs is 21\% \citep{gohil17} which is within the uncertainties of observational evidence \citep{akylas09,burlon11,brightman12,ricci17}. This consistency leads to the conclusion that the starburst discs can obscure the AGNs and may also be the dominant contributors to the peak of CXB spectrum. \footnote{The 2D NSD model was updated with energy input from AGN irradiation by following the similar scheme as described in Appendix (d) of \cite{hubeny90}. The AGN irradiation has an insignificant effect on the structure of NSDs since the energy content from the starburst outshines the AGN spectrum at parsec/sub-parsec scale while the AGN spectrum weakens at a larger distance. This result is consistent with the conclusion of \cite{gohil17} which was based on the rough estimate of the dust sublimation radius and the optical depth along the line of sight. Outflows driven by star-formation or AGN irradiation are not considered by this model.}


\section{Methodology: Evolution of the $N_{\text{H}}$ Distribution}
\label{sect:method}
\cite{gohil17} computed the $N_{\text{H}}$ distribution based on the random distribution of input parameters. In order to evolve the $N_{\text{H}}$ distribution with redshift, we take the statistical approach where the observationally motivated distribution functions of input parameters are adopted. Since these functions have a redshift dependency, one can compute the $N_{\text{H}}$ distribution with respect to redshift. Unfortunately, there is not enough observational data on the disc size $\Rout$ and the Mach number $m$. Therefore, their weights (probability) are assumed to be uniform. For $\Rout$ and $m$, we choose
\begin{align}
&R_j\equiv\Rout\in[60,120,180,240]\text{pc} \, \, \, \, \text{and}\\
&m_l\in[0.1,0.3,0.5].
\end{align}
Then, their respective weights are
\begin{align}
&W_R(z,R)=1/4, \, \, \, \,  \text{and}\\
&W_m(z,m)=1/3.
\end{align}
However, there is enough observational data which can be used to compute the distribution functions of active black hole mass $\Mbh$ and the gas fraction in galaxies.

\subsection{Evolution of the AGN Mass Function $\PhiAGN$}
Mass of a black hole plays an important role in deciding the location of starburst phenomenon in NSDs \citep{ballantyne08,gohil17}. For instance, gas has to accrete more to smaller radii before reaching the sublimation temperature of dust for a lower $\Mbh$. A bolometric luminosity of an AGN ($L$) is directly linked to a central black hole ($\Mbh$) through the Eddington radio ($\lambda=L/\Ledd$), where $\Ledd=l\Mbh$ with $l\approx1.26\times10^{38}M_{\astrosun}^{-1}$erg s$^{-1}$.
Given the Eddington ratio distribution $P(\lambda)$ and the AGN luminosity function $\Phi_L$, the AGN mass function ($\PhiAGN$) is computed as
\begin{align}
\PhiAGN(z,L)=\int_{\log(\lambda_{\min})}^{0}\Phi_L(z,L)P(\lambda)d\log\lambda.
\end{align}
In terms of $\Mbh$, $\PhiAGN$ becomes
\begin{align}
\PhiAGN(z,\Mbh)=\int_{\log[\lambda_{\min}(\Mbh)]}^{0}\Phi_L(z,L(\Mbh,\lambda))P(\lambda)d\log\lambda.
\label{eqn:phiagn}
\end{align}
The lower limit on the Eddington ratio $\lambda_{\text{min}}$ is set by the definition of an AGN ($L\ge10^{42}$ erg s$^{-1}$)\footnote{X-ray binaries with X-ray luminosity up to $10^{41.5}$\ergpers could be present in galaxies \citep{sazonov16}. Therefore, we define an AGN with the bolometric luminosity higher than $10^{42}$ \ergpers.}. Then, $\lambda_{\text{min}}$ is linked to the active black hole mass $\Mbh$ by
\begin{align}
\lambda_{\text{min}}(\Mbh)=\frac{10^{42}\ \ \text{erg s}^{-1}}{l(\Mbh/M_{\astrosun})}.
\end{align}
For the AGN luminosity function, we adopt the work of \cite{ueda14}. The bolometric luminosity ($L$) is converted into the X-ray luminosity($L_x$) by
\begin{align}
\log(L/L_x)=1.54+0.24\zeta+0.012\zeta^2-0.0015\zeta^3
\end{align}
where $\zeta=\log(L/L_{\astrosun})-12$ and $L_{\astrosun}=4\times10^{33}$ ergs s$^{-1}$ \citep{marconi04}. Unfortunately, the Eddington ratio distribution $P(\lambda)$ is very poorly known observationally. They are observed to be in two forms: log-normal and power-law distribution \citep[e.g.,][]{kollmeier06,aird12,kelly13,shankar13}. We tried both forms of the Eddington ratio distribution from \cite{tucci17} which also has dependency on redshift. The log-normal distribution from \cite{tucci17} takes the form of
\begin{align}
P(\lambda,z) =\frac{1}{2\pi\sigma(z)\lambda}\exp\Big[-\frac{(\ln\lambda-\ln\lambda_0)^2}{2\sigma^2(z)}\Big].
\label{eqn:lognormal}
\end{align}
Here, the the central value $\lambda_0$ and the dispersion $\sigma$ are
\begin{align}
\log \lambda_0(z)&=\max[-1.9+0.45z,\log(0.03)] \ \ \text{and}\ \ \\
\sigma(z) &=\max[1.03-0.15z,0.6],
\end{align}
respectively. The log-normal distribution is more favored for high luminous Type-1 AGNs and they may be related to the thin accretion flow \citep{trump11}. The power-law distribution of the Eddington ratio takes the form of \citep{tucci17}
\begin{align}
P(\lambda,z)=P_0(z)\lambda^{\alpha_{\lambda}(z)}e^{-\lambda/\lambda_0}
\label{eqn:powerlaw}
\end{align}
where
\begin{equation}\label{ch:five:sec:5:eq4:1}
\alpha_\lambda=
\begin{cases}
	-0.6 \ \ \ \ & z\le0.6\\
	-0.6/(0.4+z)\ \ \ \ & z>0.6.
   \end{cases}
\end{equation}
Here, $\lambda_0$ is 1.5 for $\eta\le0.1$ or $\lambda_0=2.5$ otherwise. Eq.~\ref{eqn:powerlaw} seems to be associated with the Type-2 AGNs \citep{aird12} and is also consistent with work of \cite{hopkins09}, \cite{kauffmann09}, and \cite{aird12} at low redshift.

Finally, one can compute the active black hole mass function by using Eq.~\ref{eqn:phiagn}. We choose a total of 16 bins in the domain of $x=\log(\Mbh)$,
\begin{align}
\begin{split}
x_i\in &[6.5,6.6,6.7,6.8,6.9,7.0,7.1,7.2,\\
&7.3,7.4,7.5,7.6,7.7,7.8,7.9,8.0].
\end{split}
\end{align}
The $\PhiAGN$ is limited to $10^8 M_{\astrosun}$ for two reasons: (1) the black hole mass function drops rapidly beyond $10^8 M_{\astrosun}$ \citep{greene06,schulze10,tucci17} and (2) we find that the $\PhiAGN$ has a very small effect on the evolution of the $N_{\text{H}}$ distribution since the evolution is mainly governed by the distribution function of $\fgout$. Then, the proper weights (probability) of each bin $W_x$ is computed by
\begin{align}
W_x(z,x)=\frac{\PhiAGN(z,x)}{\sum\limits_{i}\PhiAGN(z,x_i)}.
\label{eqn:blackholemass}
\end{align}
The left panel of Fig.~\ref{fig:wx_z} illustrates the weights of an active black hole mass $W_x$ as a function of redshift.  The solid curve represents the log-normal distribution of $P(\lambda)$ (Eq.~\ref{eqn:lognormal}), while the dashed curve is computed using the power-law distribution (Eq.~\ref{eqn:powerlaw}). The red, green, and blue color show $W_x$ for the black hole mass of $10^{6.5}M_{\astrosun}$,$10^{7.3}M_{\astrosun}$, and $10^{8.0}M_{\astrosun}$, respectively. In the case of log-normal distribution, black holes with the mass near $10^{7.3}M_{\astrosun}$ dominates in the local universe, while the lower black hole mass dominates at higher redshift. At $z\sim 2$, black holes in the entire domain of $\Mbh$ seem to have equal contribution. On other hand, when the power-law distribution is chosen, the weights of higher black hole mass dominate redshift. With either choice of $P(\lambda)$, the result shows that the weights of lower black hole mass increases with redshift and the reverse is true for the higher black hole mass. There are some differences in active black hole mass function and its evolution depending on the choice of $P(\lambda)$. However, a choice of of $P(\lambda)$ has negligible effect on the final results (the evolution of $N_{\text{H}}$ distribution and the CXB spectrum) and that is because the statistical evolution of $N_{\text{H}}$ distribution is mainly governed by the weights of $\fgout$ rather than $\Mbh$. Therefore, we select the power-law distribution $P(\lambda)$ for the further work since that particular distribution $P(\lambda)$ seems to be associated with Type-2 AGNs \citep{aird12}.

\begin{figure*}
\centerline{
\includegraphics[width=0.45\textwidth]{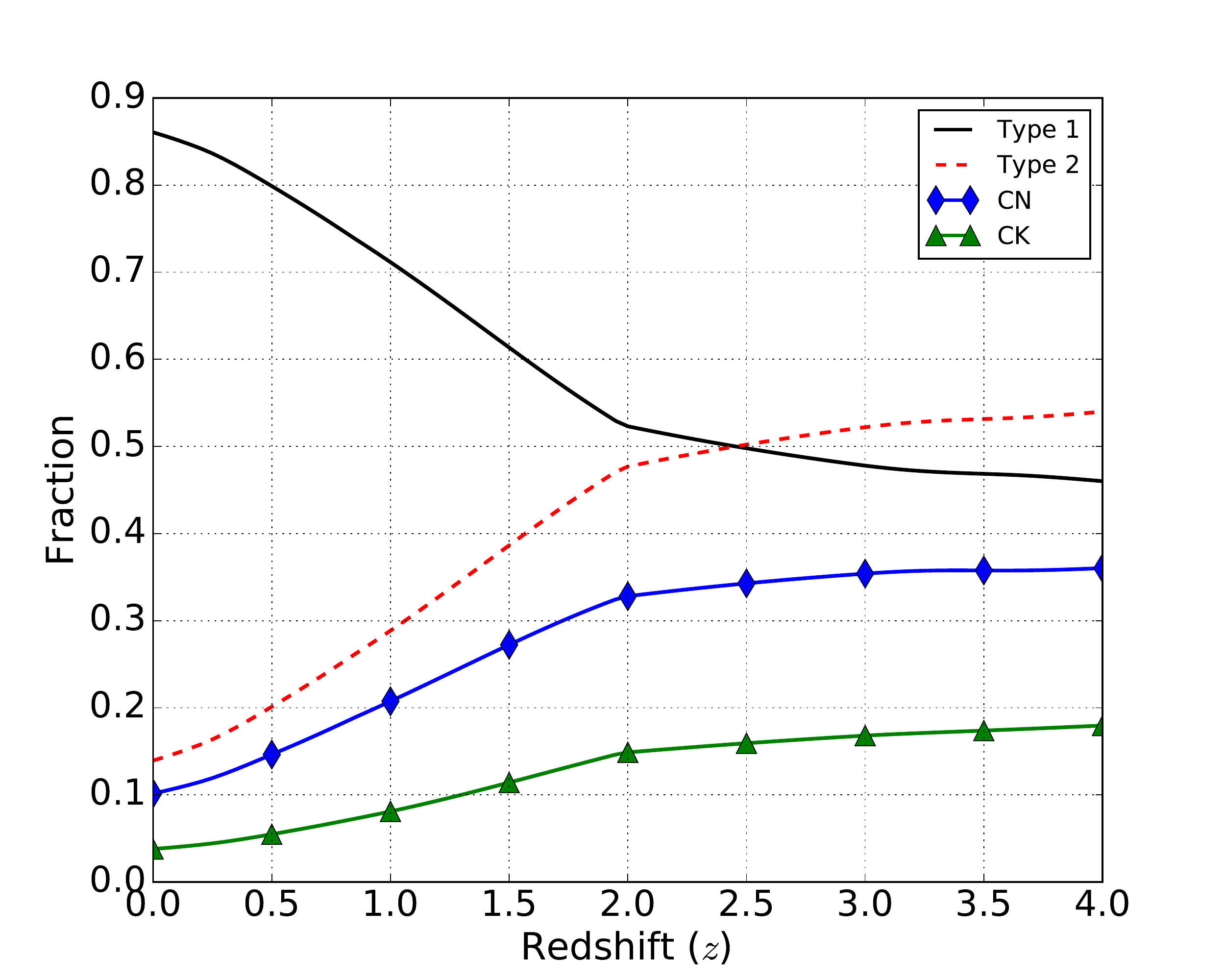}
\includegraphics[width=0.45\textwidth]{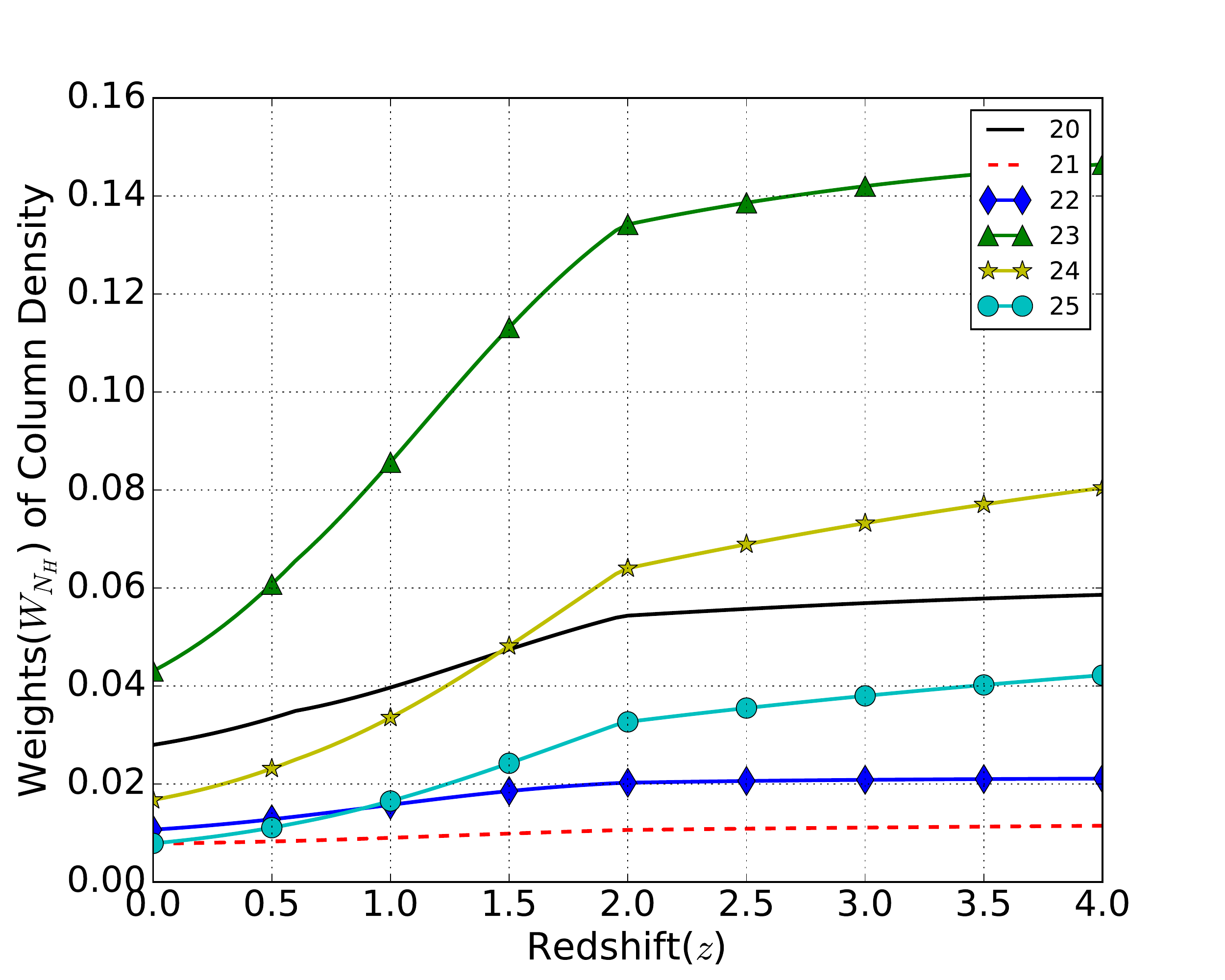}
}
\caption{\textit{Left:} The figure shows the evolution for different type of AGN fractions. The fraction of Type-1 AGNs decreases with redshift while the converse is true for the Type-2 AGNs. The fraction of Type-2 AGN, CN, and CK AGNs evolve as $(1+z)^{\delta}$ with $\delta$ equal 1.2, 1.12, and 1.45 for $z<2$, respectively. Beyond this redshift, their evolution strength decrease to 0.24, 0.16, and 0.42, respectively. \textit{Right:} The evolution of a few columns of gas (Eq.~\ref{eqn:evolve}) are shown here. All $N_{\text{H}}$ bins exhibit the evolution with $z$ and the $N_{\text{H}}\sim$ 10\tsup{23} cm\tsup{-2} dominates the distribution for all redshifts.}
\label{fig:evolve_fraction}
\end{figure*}
\subsection{Evolution of the Gas Fraction Function $\Phi_f$}
\cite{gohil17} showed that the AGN obscuration phenomenon depends on the gas fraction in NSDs. With the higher gas fraction, more amount of gas is available to accrete which increases a column of gas in annulus $\Sigmamp$ and, in turn, controls an expansion of an atmosphere. An overall gas fraction ($f_0$) in galaxies is observed to be increasing with redshift. $f_0$ is related to the depletion time and the specific star-formation rate (sSFR) through
\begin{align}
f_0(z)=\frac{1}{1+(t_{\text{dep}}(z)\text{sSFR}(z))^{-1}}.
\end{align}
The depletion time is estimated to be \citep{saintonge13}
\begin{align}
t_{\text{dep}}=1.5(1+z)^{\alpha} [\text{Gyr}]
\end{align}
with $\alpha=-1.0$ \citep{tacconi13}. \cite{lilly13} provides the analytical expression for the evolution of sSFR given by
\[
\text{sSFR}(M_*,z)=
\begin{cases}
0.07\Big(\frac{M_*}{10^{10.5}M_{\astrosun}}\Big)^{-0.1} (1+z)^3\ \ & z<2\\ \addtag
0.30\Big(\frac{M_*}{10^{10.5}M_{\astrosun}}\Big)^{-0.1} (1+z)^{5/3}\ \ & z\ge2
\end{cases}
\label{eqn:ssfr}
\]
which was motivated from observational data \citep[e.g.,][]{daddi07,elbaz07,noeske07,pannella09,stark13}.
Since there is a weak dependence on the galaxy mass $M_*$, we choose $M_*=10^{10.5}M_{\astrosun}$. Then, the distribution of gas fraction ($\Phi_f$) is assumed to be Gaussian around $f_0$ which is given by
\begin{align}
\Phi_f(z,f)=\frac{1}{\sqrt{2\pi}\sigma_f}\exp\Big[-\frac{(f-f_0)^2}{2\sigma_f^2}\Big].
\label{eqn:phif}
\end{align}
The dispersion $\sigma_f$ is estimated to be 0.2 by roughly fitting the Gaussian to the results of \cite{tacconi13}. Finally, we can compute the weights $W_f$ of input parameter $f\equiv\fgout$ from
\begin{align}
W_f(z,f)=\frac{\Phi_f(z,f)}{\sum\limits_{k}\Phi_f(z,f_k)}
\label{eqn:gasfraction}
\end{align}
where
\begin{align}
f_k\in[0.2,0.4,0.6,0.8].
\end{align}
The right panel of Fig.~\ref{fig:wx_z} shows the evolution of 20\%, 40\%, 60\%, and 80\% gas fraction with redshift which are represented by black, red, blue, and green curves, respectively. At low redshift, NSDs with low gas fraction dominates the sample, while NSDs with a 40\% gas fraction dominates at 1<$z$<2. Beyond $z=2$, NSDs with a 60\% gas fraction dominates. $W_f$ of NSDs with low gas fraction decreases overall with redshift and the reverse is true for the the NSDs with the high gas fraction (60\% and 80\%). Similar to the observational evidence, the figure illustrates that the dominant gas fraction increases with $z$ and its evolution flattens out at higher redshift.

\subsection{Evolution of the $N_{\text{H}}$ Distribution}
Using redshift dependent weights of input parameters ($W_x(z)$,$W_R$,$W_f(z)$, and $W_m$), one can evolve the $N_{\text{H}}$ distribution with redshift $W_{N_{\text{H}}}(z)$. The $N_\text{H}$ distribution is divided among 11 bins
\begin{align}
\begin{split}
\log[N_H(\percmsqr)]\in&[20.0,20.5,21.0,21.5,22.0,22.5,\\
&23.0,23.5,24.0,24.5,25].
\end{split}
\end{align}
For a given NSD disc, a viewing angle (an orientation of an NSD with respect to an observer) is divided into 30 bins between 0 and 90 degrees. Then, the weights of $N_{\text{H}}$ bins  $W_{N_{\text{H}}}(\vec{I})$ for the disc is given by
\begin{align}
W_{N_{\text{H}}}(\vec{I}_P,N_{\text{H}}) =N(N_{\text{H}})/N_{\text{tot}}
\end{align}
where $N(N_{\text{H}})$ is the number of column density with $N_{\text{H}}$ and $N_{\text{tot}}$ is the total number of columns of gas. The weights of $\log[N_H(\percmsqr)]=20.0$ and $\log[N_H(\percmsqr)]=25.0$ bins include all the lines of sight with column density $N_H\le 10^{20}\percmsqr$ and $N_H\ge 10^{25}\percmsqr$, respectively. $\vec{I}_P$ is the input parameter vector equal to $[x,\Rout,\fgout,m]$.
In total, 768 models are computed across the input parameter space which gives 23,040 ($768\times 30$) columns of gas. Then, the evolution of $N_{\text{H}}$ distribution is given by
\begin{align}
W_{N_{\text{H}}}(z,N_{\text{H}}) =A_N\sum\limits_{i,j,k,l}W_{N_{\text{H}}}(\vec{I}_P,N_{\text{H}})\Pi(z,\vec{I}_P)
\label{eqn:evolve}
\end{align}
where
\begin{align}
\Pi(z,\vec{I}_P) &=W_x(z,x)\times W_f(z,f)\times W_m\times W_R
\end{align}
and $A_N$ is the normalization constant.
By using Eq.~\ref{eqn:evolve}, the evolution of Type-1, Type-2, CN and CK fractions can be predicted, as well as the evolution of each column density $N_{\text{H}}$. The results are shown in Fig.~\ref{fig:evolve_fraction}.

\section{Results}
\label{sect:result}
By employing an adequate statistical method, we compute the evolution of AGN obscuration. By utilizing this result, we later predict the cosmic X-ray background and the AGN number counts in 2-8 keV and 8-24 keV bands as well as in different $N_{\text{H}}$ bins.
\subsection{Evolution of AGN obscuration}
A simple statistical evolution shows that obscuring material has a strong evolution with redshift which is illustrated in Fig.~\ref{fig:evolve_fraction}. The left panel exhibits that the evolution of Type-1, Type-2, CN, and CK AGN fractions are represented by black-solid, red-dashed, blue-diamond, and green-triangle curves, respectively. All the fractions of obscured AGNs increase with redshift, while the Type-1 AGN fraction decreases.\footnote{The sample of $N_{\text{H}}$ distribution has a column of gas along all the line of sight from $0^{\circ}$ to $90^{\circ}$. We added the galactic scale obscuration ($10^{20}$ cm$^{-2}$) to the line of sights which has zero column density from nuclear regions.}  The AGN fractions have a strong evolution up to $z=2$ which is also concluded by many other groups \citep{ballantyne06,treister06,ueda14,liu17}. For $z>2$, there is a weak dependence on redshift which is also consistent with the work of \cite{liu17}. The drastic change at $z=2$ can be explained by the discontinuous sSFR function which is given by Eq.~\ref{eqn:ssfr}. The ratio of CN AGN fraction ($\fCN$) to $\fCK$ is always higher than 1 throughout the redshift range. The right panel shows the evolution of a few individual column density bins. The black-solid, red-dashed, blue-diamond, green-triangle, yellow-star, and cyan-circle curves represent a column of gas with $\log[N_{\text{H}}(\text{cm}^{-2})]$ equal 20, 21, 22, 23, 24, and 25, respectively. In general, their weights $W_{N_{\text{H}}}$ are predicted to increase with redshift. The panel exhibits that the $N_{\text{H}}$ distribution peaks near 10\tsup{23} cm\tsup{-2} throughout the redshift range which is in agreement with observations \citep[e.g.,][]{burlon11,ueda14,buchner15,sazonov15}. The unobscured AGNs with $N_{\text{H}}=10^{20}$ cm$^{-2}$ have a strong evolution while the AGNs with $N_{\text{H}}=10^{21}$ cm$^{-2}$ possess a weak dependence on $z$.

\begin{figure}
\includegraphics[clip,width=\columnwidth]{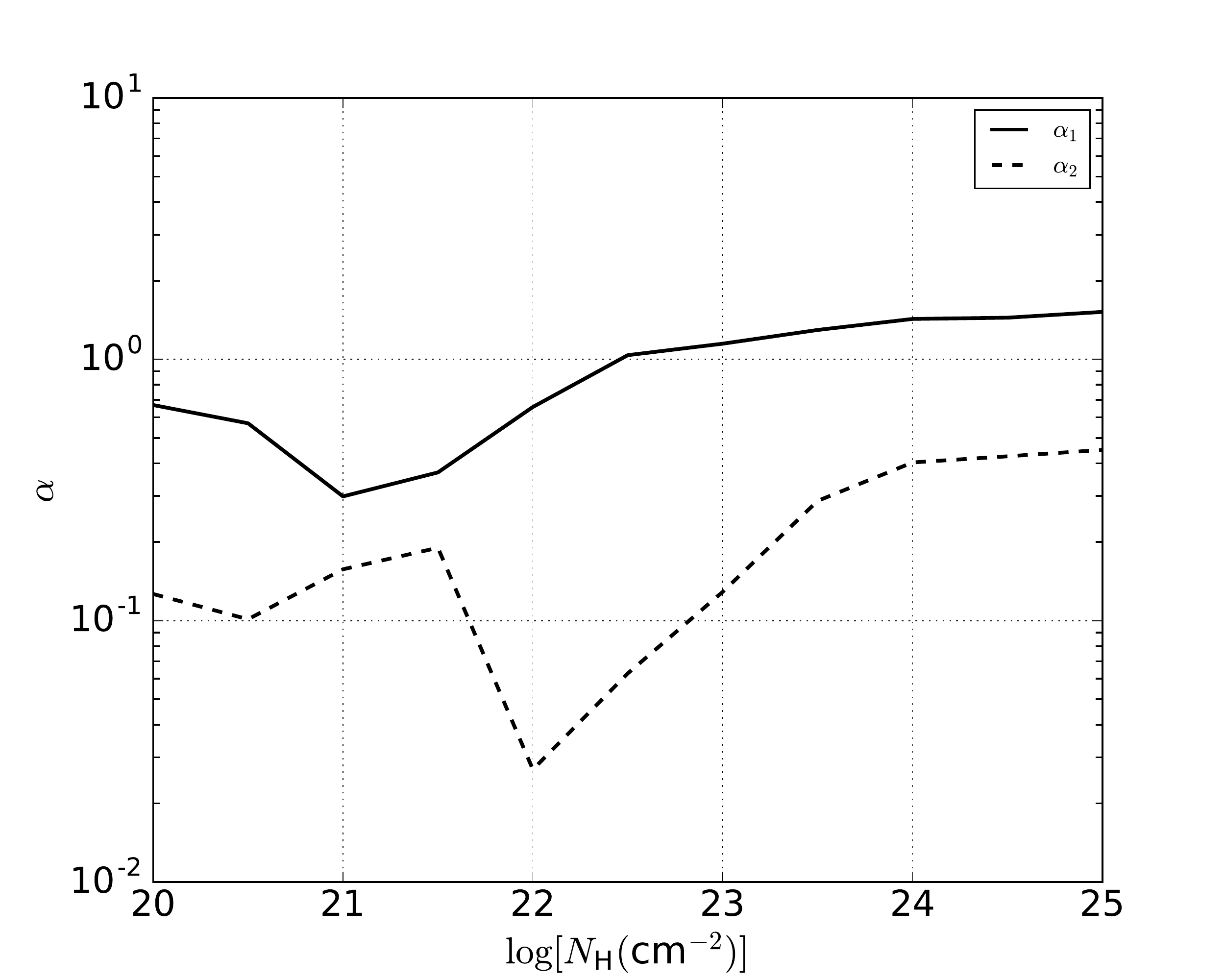}
\caption{Figure illustrates the dependence of evolution strength $\alpha$ on $N_{\text{H}}$, where $\alpha$ is the power-law index such that $W_{N_\text{H}}\propto (1+z)^{\alpha}$. AGNs with increasing $N_{\text{H}}$ have a faster evolution for both cases: $z<2$ (solid curve) and $z>2$ (dashed curve). In other words, the difference between the fraction of a given $N_{\text{H}}$ bin at $z=0$ and $z=2$ significantly rises as the $N_{\text{H}}$ increases.}
\label{fig:alphas}
\end{figure}

\begin{table}
\centering
\caption{The redshift evolution $(1+z)^{\delta}$ of Type-1, Type-2, CN, and CK AGN fractions are characterized by the power-law index $\delta$. $\delta=\delta_1$ and $\delta=\delta_2$ represent the evolution strength for $z<2$ and $z>2$, respectively.}
\label{table:delta}
\begin{tabular}{c | c c c c}
Obscuration & $\delta_1 (z<2)$ & $\delta_2 (z>2)$& \\\hline
$f_1$     & -0.43  & -0.28  &  \\
$f_2$       & 1.2  & 0.24     &  \\
$\fCN$     & 1.12  & 0.16    &  \\
$\fCK$     & 1.45   & 0.42   &
\end{tabular}
\end{table}

The evolution strength of the AGN fraction can be measured by fitting the power-law $(1+z)^{\delta}$. $\delta=\delta_1$ and $\delta=\delta_2$ represent the power-law index for $z<2$ and $z>2$, respectively and their values are shown in Table~\ref{table:delta}. The $\fCK$ always has a slightly stronger evolution than $\fCN$. The strength of an evolution for each $N_{\text{H}}$ bin is again measured by fitting the power-law $W_{N_{\text{H}}}\propto(1+z)^{\alpha}$. The power index $\alpha=\alpha_1$ for $z<2$ (solid curve) and $\alpha=\alpha_2$ for $z>2$ (dash curve) as a function of a column density are illustrated in Fig.~\ref{fig:alphas}. The power-law indices are predicted to increase with the amount of obscuration for obscured AGNs in both eras: $z<2$ and $z>2$. The $\alpha$ has a weak dependence on column density in Compton thick regime. Fig.~\ref{fig:alphas} also suggests that the evolution of AGN environment highly depends on obscuring medium. The difference between the AGN fraction of a given $N_{\text{H}}$ bin in the local universe and at $z=2$ increases as the amount of obscuration ($N_{\text{H}}$) increases. Similar conclusion holds true while comparing the AGN fraction at $z=2$ and $z=4$.

\begin{figure}
\centerline{
\includegraphics[width=0.5\textwidth]{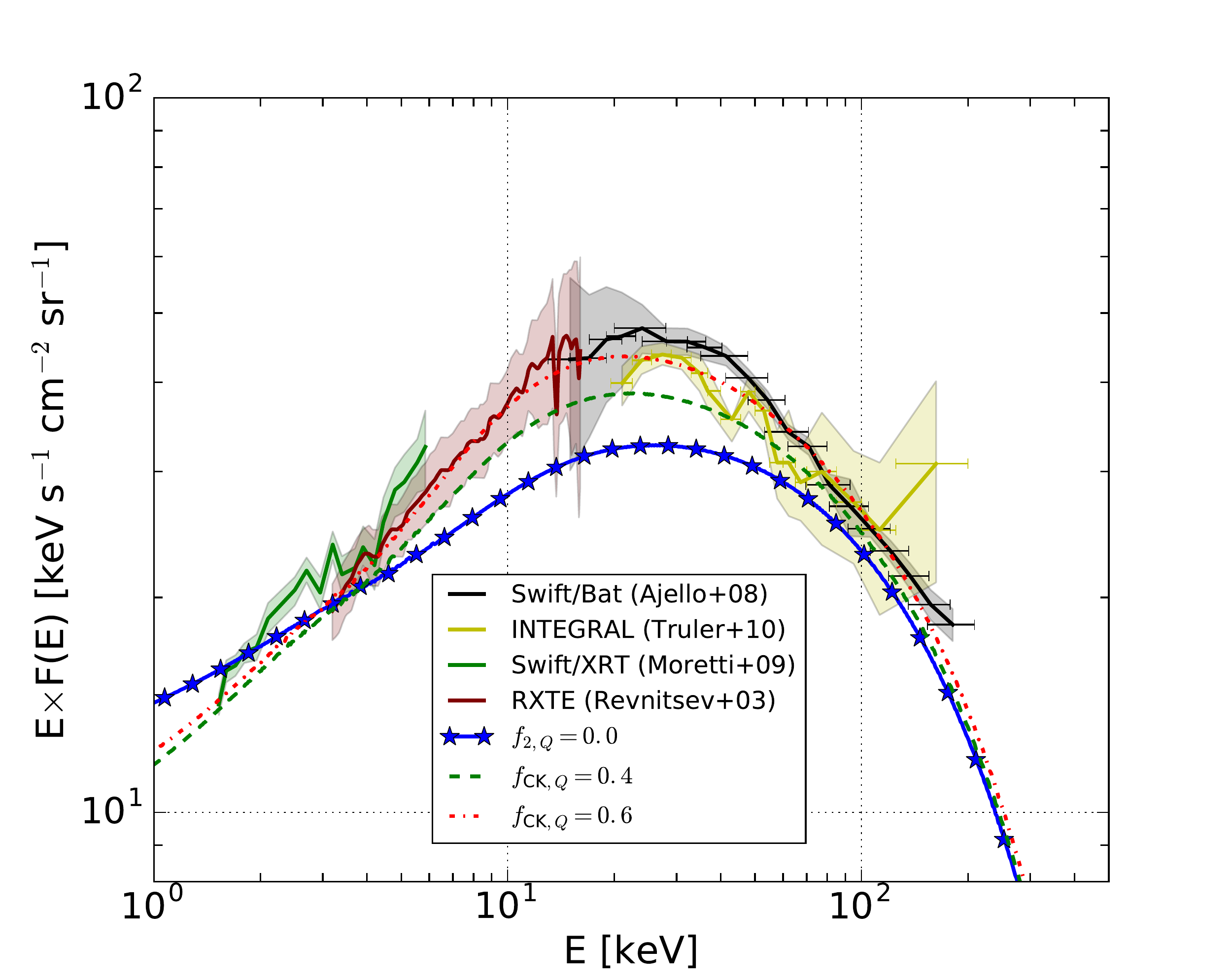}
}
\caption{The predicted CXB spectra are shown by assuming the photon-index $\langle\Gamma\rangle=1.85$ \citep{ueda14} for an intrinsic AGN X-ray spectrum and the high energy cutoff $E_c=$ 220 keV. The shaded area represents uncertainty in observational data. The blue-star curve is the predicted CXB spectrum produced by AGNs with $L_{2-10\text{keV}} <10^{44}$ erg s$^{-1}$ (Seyfert regime). These AGNs are obscured by nuclear starburst discs with the $N_{\text{H}}$ distribution given by the 2D NSD theory while the AGNs with $L_{2-10\text{keV}} >10^{44}$ erg s$^{-1}$ (quasar regime) are assumed to be unobscured. The $E<3.0$ part of the spectrum is overestimated while the spectrum has a lower peak (blue-star curve) with comparison to the observed one (shaded area). In order to produce the correct SED, at least 40 per cent of CK ``quasars-like'' AGNs are required (green-dashed curve). The contribution of 20 per cent CN and 60 per cent CK quasars predict the best fit spectrum to observations which is shown by the red-dash-dotted curve. The spectrum has $\Gamma=1.49$ in 2-10 keV bands which is consistent with observations \citep[e.g.,][]{moretti09,cappelluti17}}.
\label{fig:cxb_result}
\end{figure}

\begin{figure}
\centerline{
\includegraphics[width=0.5\textwidth]{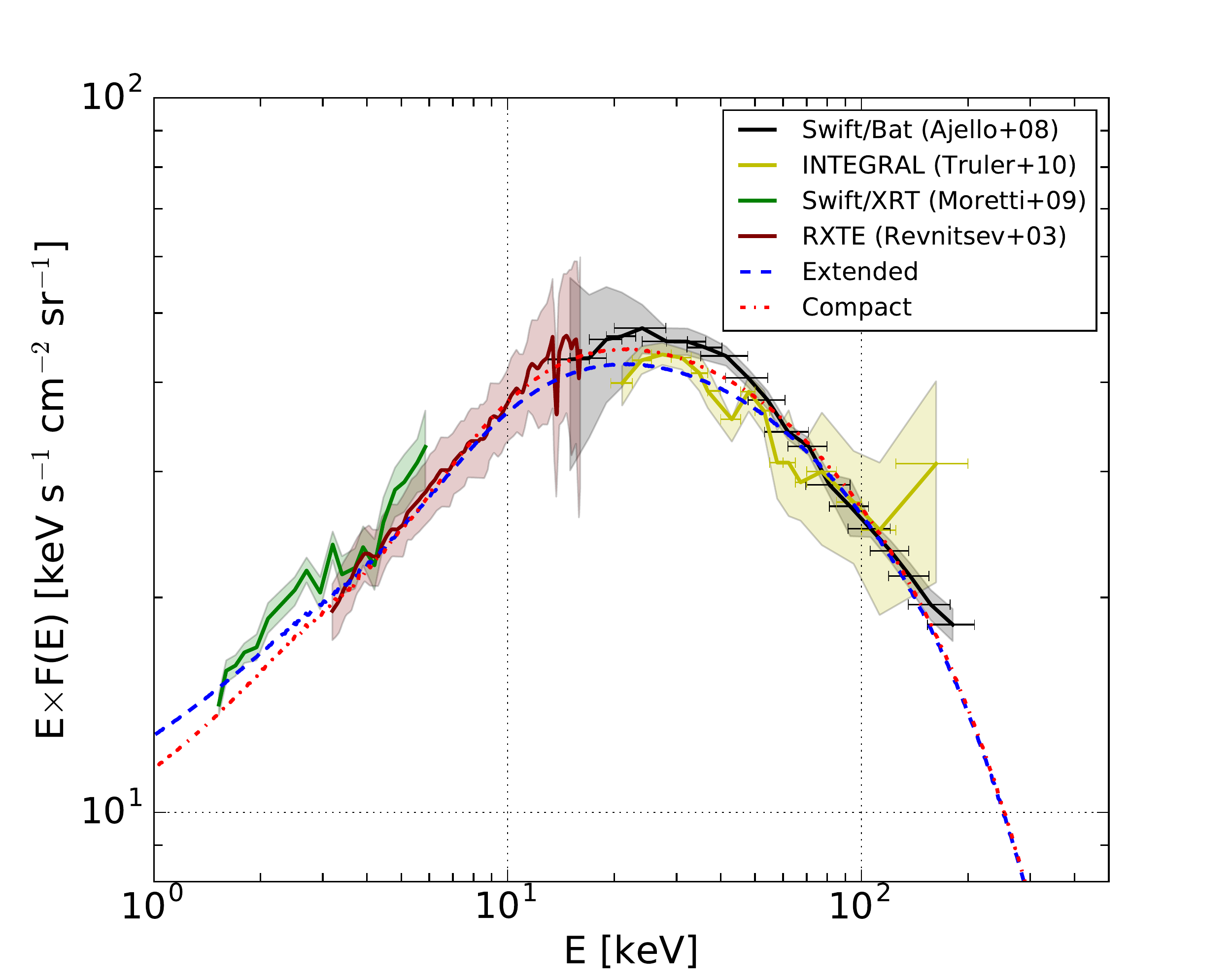}
}
\caption{The compact ($\Rout$ with 60 pc and 120 pc) and extended ($\Rout$ with 180 pc and 240 pc) NSDs do not have a significant difference on the CXB spectrum. However, the peak of the spectrum has a slightly better match with the observations when only the compact NSDs are used. The 2-10 keV photon index ($\Gamma$) is 1.46 in the case of compact NSDs while the case of extended NSDs predict it to be 1.52.}
\label{fig:cxb_analysis}
\end{figure}

\begin{figure*}
\centerline{
\includegraphics[width=0.5\textwidth]{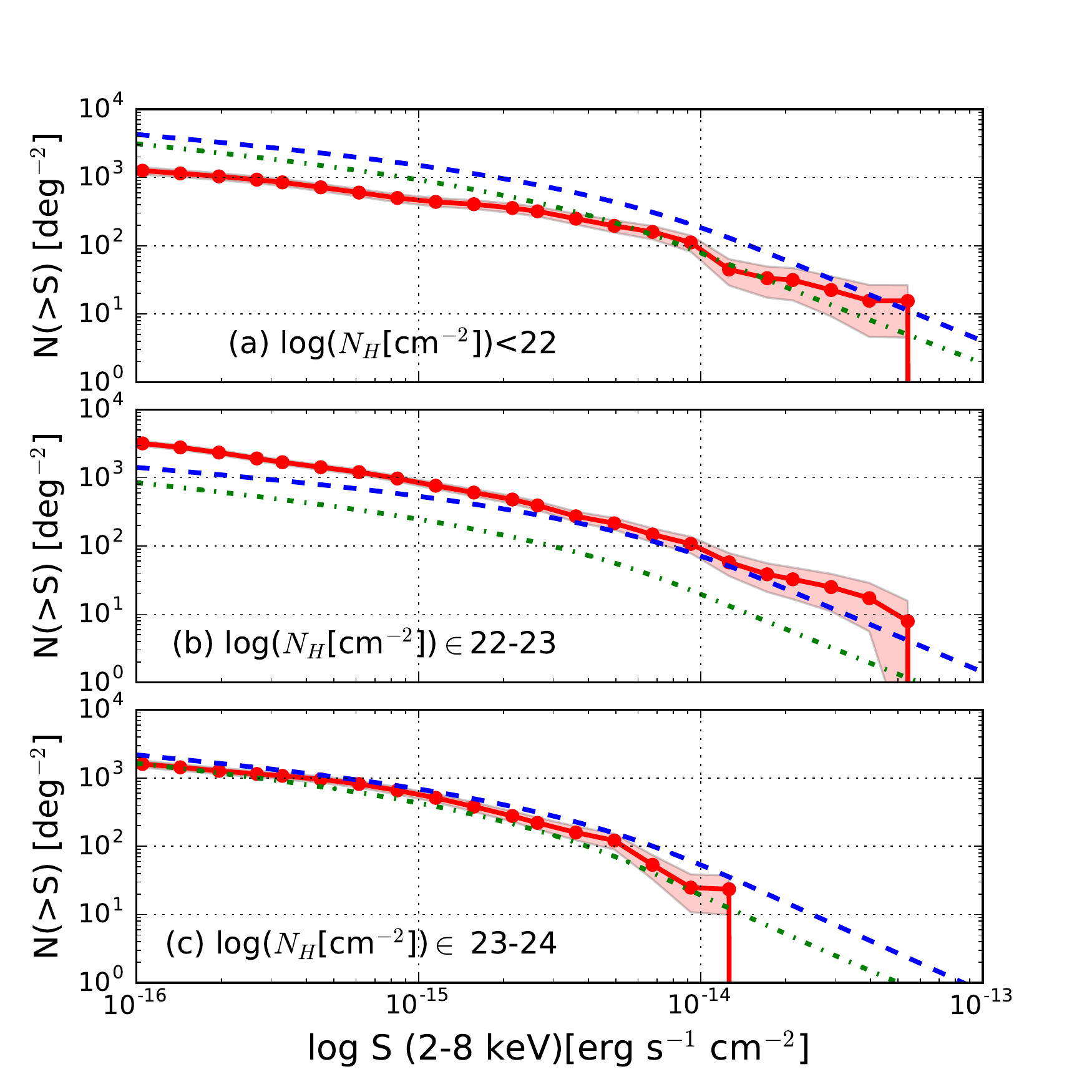}
\includegraphics[width=0.5\textwidth]{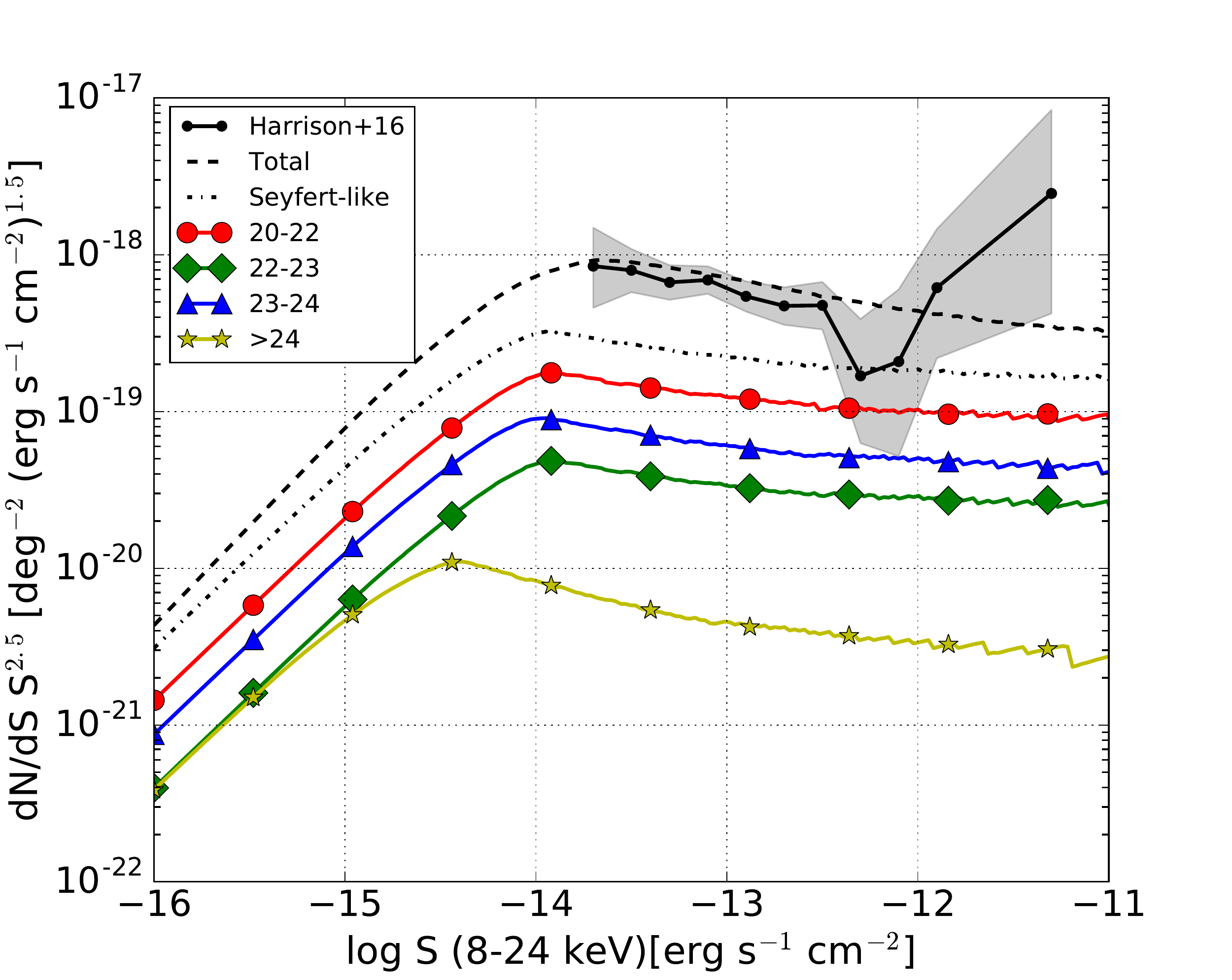}
}
\caption{\textit{Left: }Figure illustrates the number counts ($N$) of unobscured, moderately obscured, and heavily obscured AGNs in 2-8 keV bands through panel (a), (b), and (c) respectively. The number counts of moderately obscured AGNs at high energy flux ($S$) and the heavily obscured AGNs are in fair agreement with the observational data from CDF-S \citep{lehmer12}. The green-dash-solid shows the number counts of ``Seyfert-like'' AGNs which are obscured by the $N_{\text{H}}$ distributions associated with NSDs. The blue-dashed curve shows the total number counts which also includes the contributions from quasars (20 per cent CN and 60 per cent CK). \textit{Right: } The total differential AGN number counts in 8-24 keV band (black-dashed curve) are shown in this panel which agrees with the observational sample of \textit{NuSTAR} \citep{harrison16}. The panel also provides the prediction of differential number counts of AGNs in many obscuration bands associated with the ``Seyfert-like'' AGNs. Their total differential number counts are shown with the black-dash-dotted curve. The figure also illustrates that contribution from quasars is important in order to match the observations.}
\label{fig:numbercounts}
\end{figure*}

\subsection{Cosmic X-ray Background}
By utilizing redshift dependent $N_{\text{H}}$ distribution $\WNH(N_{\text{H}},z)$ and starting with the work of \cite{ballantyne11}, we predict the cosmic X-ray background (CXB) associated with the dusty starburst regions.\footnote{The AGN luminosity function used is from the work of \cite{ueda14}.} Since the average photon index of an intrinsic AGN X-ray spectrum $\langle\Gamma\rangle$ is observed to peak near 1.85 \citep{ueda14}, the $\langle\Gamma\rangle$ is fixed at 1.85. The high energy cutoff $E_c=220$ keV is chosen such that the best fit to the observed CXB spectrum is produced. We also introduce the redshift dependent reflection fraction $R_f(z)$ which is defined as
\begin{align*}
R_f(z)\equiv \frac{4\pi f_{2,R_f}(z)}{2\pi}.
\end{align*}
Here, the fraction of AGNs with high reflection $f_{2,R_f}(z)$ is computed using the $N_{\text{H}}(z)$ distribution from the 2D NSD theory. Since, high reflection is also observed from moderately obscured CT AGNs \citep{ricci11,esposito16}, $f_{2,R_f}(z)$ is defined as the fraction of $N_{\text{H}}$ from 10$^{23.5}$ to 10$^{25}$ cm$^{-2}$. The $N_{\text{H}}$ distribution of NSDs is used for $L_{2-10\text{keV}} <10^{44}$ erg s$^{-1}$ (Seyfert regime) since they are more viable source of obscuration for ``Seyfert-like'' AGNs \citep{ballantyne08,gohil17}. In the case of high luminosity AGNs, very efficient radial transport of gas (i.e. high Mach number) is required. However, the gas is fueled more into star-formation rather than to a central black hole as the Mach number increases \citep{ballantyne08,gohil17}. Hence, NSDs are less likely to survive in quasars within the scope of physics which is presented in the 2D NSDs theory. Therefore, for $L_{2-10\text{keV}} >10^{44}$ erg s$^{-1}$ (quasar regime), we adopt two cases: (i) only unobscured AGNs in order to study the consequences of solely NSDs on CXB spectrum, and (ii) the fraction of Type-1 (T1), CN, and CK such that it produces the best fit to the observed CXB spectrum.

The predicted CXB spectra are shown in Fig.~\ref{fig:cxb_result}. The shaded regions are the uncertainty in the observed CXB spectrum. The blue-star curve represents the first case without any contributions of obscured quasars ($f_{2,Q}=0.0$). The high energy part of spectrum matches the observation which is due to the selection of a particular energy cutoff $E_c$. The spectrum is predicted lower in general, but the converse is true for $E<3.0$ keV. Straightaway, we can find the required contribution of ``quasar-like'' AGNs in order to predict the observed CXB spectrum. The green-dashed curve shows that at least 40 per cent of such CK AGNs ($f_{\text{CK},Q}$) along with 0.4 and 0.2 fractions of CN and T1 ``quasars-like'' AGNs, respectively are required in order to produce the spectrum within observed uncertainty. The 0.2, 0.2, and 0.6 fractions of T1, CN, and CK yields the best matched spectrum which fits very well with the observed one. The 2-10 keV photon index is 1.49 which is consistent with the observed value, 1.4-1.54 \citep{marshall80,deluca04,moretti09,cappelluti17}.

We also explore the effect of compact and extended NSDs on CXB spectrum which are illustrated by red-dash-dotted and blue-dashed curves in Fig.~\ref{fig:cxb_analysis}, respectively. The NSDs with $\Rout$ equal 60 pc and 120 pc are considered as compact, while the extended NSDs are referred to the one with the size of 180 pc and 240 pc. A sample with either type of NSDs produces the CXB peak (20-30 keV) within the observational uncertainty; however, the NSDs with a smaller size are more favorable. The compact and extensive NSDs predict the photon index of 2-10 keV to be 1.46 and 1.52, respectively which are still in good agreement with observations \citep{marshall80,deluca04,moretti09,cappelluti17}.

\subsection{AGN Number Counts}
A widely used physical quantity in studying the cosmic X-ray background is the AGN number counts per unit area $N$ \citep[e.g.,][]{brandt01,bauer04,kim07,georgakakis08,lehmer12,harrison16}. Therefore, we predict the AGN number counts in 2-8 keV and 8-24 keV bands for the case of best matched CXB spectrum with the observational one (i.e., $f_{2,Q}=0.8$). Afterward, they are compared against the observations made by 4 Ms \textit{Chandra} Deep Field-South (CDF-S) survey and \textit{NuSTAR}. The left panel of Fig.~\ref{fig:numbercounts} shows the predicted AGN number counts in 2-8 keV as a function of energy flux ($S$) where panels (a), (b), and (c) are sorted by a range of column density. The red curve represents the observed number counts and the shaded area is observational uncertainty \citep{lehmer12}. The AGN number counts with $L_{2-10\text{keV}} <10^{44}$ erg s$^{-1}$ are shown in green-dash-dotted curve and the blue-dashed show the total number of AGN number counts. The number counts of unobscured AGNs (top-left panel) is overestimated even with the exclusion contribution from quasars. The obscured AGN number counts in $22\le\log(N_{\text{H}}[\text{cm}^{-2}])<23$ range is underestimated. With the inclusion of quasars contribution, the number counts fits the observed ones for high energy flux, but it is still lower for $S<10^{-14.7}$ erg s$^{-1}$ cm$^{-2}$. For heavily obscured AGN (10$^{23}$-10$^{24}$ cm$^{-2}$), the contribution from quasars is very small and the AGN number counts are in fair agreement with observations \citep{lehmer12} as shown in bottom-left panel of Fig.~\ref{fig:numbercounts}. This suggests that the 2D NSD theory produces the correct form of $N_\text{H}$ distribution for heavily obscured AGNs in Seyfert galaxies.

A major source for overestimating the unobscured AGNs could be an exclusion of galaxy-scale obscuration. The CXB model only includes the obscuring medium which resides within a couple of hundred parsec scale (NSDs). Inclusion of galaxy-scale obscuration would produce a better match with observations for both unobscured AGNs and weakly-obscured AGNs (middle-left panel of Fig.~\ref{fig:numbercounts}) since this will shift some of the AGNs number counts from unobscured AGNs bin (top-left panel of Fig.~\ref{fig:numbercounts}) to $22\le\log(N_{\text{H}}[\text{cm}^{-2}])<23$ bin. Other possible sources for discrepancy include cosmic variance in observational data \citep{somerville04} and a first-order estimate for $N_\text{H}$ based on energy band ratios \citep{lehmer12}.

The right panel of Fig.~\ref{fig:numbercounts} exhibits the differential number counts ($dN/dS$) in 8-24 keV band and are compared against the \textit{NuSTAR} observation \citep[e.g.,][]{harrison16}. The black-solid line represents the observational data with uncertainty in shaded area and this is in fair agreement with the predicted total differential AGN number counts (black-dashed line) which also includes the quasar contribution. The black-dash-dotted curve shows the total differential counts of only ``Seyfert-like'' AGNs (associated with NSDs) which alone are not sufficient enough to produce the observed ones. The red-circle, green-diamond, blue-triangle, and yellow-star curves represent ``Seyfert-like'' AGNs ($L_{2-10\text{keV}} <10^{44}$ erg s$^{-1}$) with the obscuration in the range of $\log(N_{\text{H}}[\text{cm}^{-2}])\in$ 20-22, 22-23, 23-24, >24, respectively. The figure suggests that the unobscured differential counts dominate the entire flux range. Furthermore, the differential counts of AGNs with 10$^{23}$-10$^{24}$ cm$^{-2}$ range dominates the sample of obscured AGNs.

\section{Discussion}
\label{sect:limitation}
Unlike previous CXB models \citep{ueda03,treister05,ballantyne06,gilli07}, we include redshift-dependent distribution of column density $W_{N_{\text{H}}}(z,N_{\text{H}})$ and the reflection parameter $R_f(z)$ motivated from a physical model of NSDs in modeling of the CXB. This CXB model comprises of a diverse evolution of CN AGNs (e.g., $\delta=1.2)$ and CK AGNs (e.g., $\delta=1.45$). However, the predicted CXB spectrum mainly due to the 2D NSD models is lower in comparison to the observed spectrum. This is not surprising for many reasons. The CXB traces an entire history of AGNs from low to high redshift. NSDs are a more viable source of the AGN obscuration at the intermediate redshift \citep{ballantyne08,gohil17}. Moreover, there are also other physical mechanisms which are not included in this modeling and they can play an important role in AGN obscuration such as the magnetic field, outflow, and extreme high Mach number flow (shock wave regions). For instance, outflows are observed in many star-forming galaxies \citep{tremonti07, heckman11, diamond12} which can reduce the amount of obscuration through removal of gas. Besides the ``Seyfert-like'' AGNs, very luminous quasars are also observed to be obscured \citep{iwasawa12,buchner15}. The obscuration in their environment may be driven by different mechanisms than the star-forming regions in ``Seyfert-like'' AGNs. The assumptions such as the evolution of gas fraction at hundreds of parsec scale being the same as the galaxy gas fraction, and the random selection of $\Rout$ and $m$ can also contribute to errors in final results.

\subsection{Comparison with observed fraction of obscured quasars}
Our work suggests that 80 per cent of the quasars need to be obscured in order to produce the observed CXB spectrum besides the $N_{\text{H}}$ distribution associated with the dusty starburst discs. This value is in fair agreement with the observational work done by many groups. \cite{buchner15} find that the cosmic averaged obscured fraction of AGNs with X-ray luminosity $L_x>10^{43}$ erg s$^{-1}$ is 0.72-0.81 using ~2000 sample size from a various surveys. \cite{iwasawa12} detect $\approx$0.75 fraction of obscured ``quasar-like'' AGNs in deep XMM-CDFS survey sample at $z>1.7$. \cite{schawinski12} estimate $\sim 90$ per cent of observed quasars in $1<z<3$ by deep HST WFC3/IR imaging to be heavily obscured. Using Chandra Deep Field
South (CDF-S) data in 4-7 keV band, \cite{wang07} find that $71\pm 19$ per cent of quasars are obscured (Type-2). Furthermore, a majority of these ``quasar-like'' AGNs are predicted to be CK. Our CXB model requires 40 (lower limit) to 60 per cent (best match) of these high luminosity AGNs to be CK. Similar conclusions are also reached by many groups \citep[e.g.,][]{martinez07,draper10}. \cite{martinez07} finds that 10 out of 12 observed quasars at $z>1.7$ are Type-2 and $\sim 67$ per cent of them are likely to be Compton-thick. Based on the analysis of Chandra and XMM-Newton data, \cite{jia13} estimate the fraction of Compton-thick in Type-2 quasars sample to be 46 to 64 per cent at $z<0.73$. By studying the sample of 33 mid-IR luminous quasars at $1<z<3$, \cite{delmoro15} find 24-48 per cent of $\fCK$ quasars which is a slightly lower in comparison with the work of above mentioned groups and our study. However, the reported Type-2 fraction of quasars by \cite{delmoro15} is 67 to 80 per cent which is impressively in fine agreement with the predicted fraction 60-80 per cent.

\subsection{Evolution of the obscured AGNs fraction}
The evolution of AGN fractions is expected to be overestimated by the NSD models due to several reasons mentioned above. Therefore, the predicted AGN fractions should be taken as a lower limit. We again emphasize that the AGN fractions are computed with respect to the entire domain of a sample, $N_{\text{H}}\in [0.0,10^{26}]$ \percmsqr; for instance,
\begin{align}
f_2\equiv\frac{N(22\leq\log[N_{\text{H}}(\percmsqr)]\leq26)}{N(20\leq\log[N_{\text{H}}(\percmsqr)]\leq26)},
\end{align}
where $N$ is the number of obscured AGNs in that given $N_{\text{H}}$ bin. The power-law evolution of obscured AGNs $f_2$ $\propto (1+z)^{\delta}$ associated with the dusty starburst discs (``Seyfert-like'' AGNs) is governed by $\delta=1.2$ for $z<2$. Many groups have predicted $\delta$ in 0.4-0.5 range \citep{ballantyne06,treister06,ueda14,liu17}. Since, the number density of ``quasar-like'' AGNs peaks at $z\sim 2-3$ \citep{ueda03, barger05,silverman08,yencho09,aird12}, $f_{2,Q}$ can play an important role at $z\sim 2$. Therefore, if the quasar contribution ($f_{2,Q}=0.8$) is added to the obscured AGN fraction at $z=2$ and one requires $\delta$ to be 0.45, then the estimated $f_2$ is approximately 40 per cent in the local universe. This serves as the lower limit since the presence of outflows at higher redshift are more likely to remove gas from a system which will reduce the amount of obscuration resulting in lower $\delta$ (<0.45).  However, the estimated $f_2(z=0)=0.4$ is lower than the work of \cite{burlon11} who finds that $\fCN$ by itself is around 50 per cent at $L\sim 10^{43.8}$\ergpers in the local universe. Moreover, the power-index of $\fCN$ evolution is observed to be steeper with $\delta=0.6$ in luminosity range of $10^{43.5}-10^{44.2}$\ergpers \citep{liu17} than the power-index of $\fCK$ AGNs ($\delta=0.45$). However, we find an opposite conclusion in scenario of AGNs associated with NSDs. In that case, the $\fCK$ has a stronger evolution with comparison to $\fCN$.

\subsection{Evolution of Compton-thick fraction and reflection parameter}
Analysis similar to the above section can also be done for the evolution of the CK fraction $\fCK$ $\propto (1+z)^{\delta}$. With the contributions from ``quasars-like'' AGNs, $f_{2,Q}$ corresponds to $\sim 37$ per cent at $z=2$. The recent studies have found 0.27-0.46 fraction of CK AGNs in the local universe \citep{buchner15,ricci15,lansbury17}. If we take the observed value of $\fCK$ in local universe to be around 35 per cent \citep{brightman12,buchner15}, then a very weak evolution, $\delta=0.05$, is predicted, which is consistent with the conclusion of \cite{brightman12} and \cite{buchner15}. However, other groups have also inferred $\fCK\sim 20$ per cent in local universe from observations \citep{akylas09,brightman11a,brightman11b,burlon11}. In that case, the $\fCK$ is predicted to evolve with $\delta\sim 0.56$. Using \textit{NuSTAR} extragalactic survey data, \cite{delmoro17} find that the reflection ($R_f$) is around 0.5 for AGNs with $\Gamma=1.8$ and \textcolor{blue}{Zappacosta et al. (submitted, 2017)} find the mean value to be $R_f=0.41$ at $0<z<2.1$. These values are in fair  agreement with our introduced definition of reflection quantity ($R_f$). The $R_f$ associated with the NSDs rises from 0.13 at $z=0$ to 0.47 at $z=2.0$ and 0.58 at $z=4.0$.

\section{Conclusion}
\label{sect:conclusion}
To summarize, we statistically evolve the $N_{\text{H}}$ distribution ($W_{N_{\text{H}}}(z,N_{\text{H}})$) using redshift dependent distribution of the input parameters: the gas fraction $\fgout$ (Eq.~\ref{eqn:phif}) and the black hole mass $\Mbh$ (Eq.~\ref{eqn:phiagn}). Then, the $N_{\text{H}}(z)$ is utilized to predict the evolution of Compton thin and CK AGNs. In addition, by utilizing the $W_{N_{\text{H}}}(z,N_{\text{H}})$ (Eq.~\ref{eqn:evolve}), we predict the CXB spectrum and the AGN number counts in 2-10 keV and 8-24 keV bands. Below we summarize our main findings:
\begin{itemize}
\item The obscured AGNs show a strong positive evolution with redshift up to $z=2$ and it is weakened afterward. The main driver of the evolution is gas-fraction in galaxies.
\item The fraction of CN and CK AGNs do not evolve $\propto (1+z)^{\delta}$ in the same manner. The predicted power-law index of evolution $\delta$ is 1.12 and 1.45 for $\fCN$ and $\fCK$ for $z<2.0$ which are associated solely with the dusty starburst regions. Moreover, the ratio of $\fCK$ to $\fCN$ is always less than 1.
\item Within the sample of obscured AGNs, the distribution column density along the line of sight peaks near 10\tsup{23} cm\tsup{-2} independently of redshift \citep[e.g.,][]{burlon11,ueda14,buchner15,sazonov15}.
\item Based on the \textit{NuSTAR} extragalactic survey sample, \cite{delmoro17} infer the reflection parameter $R_f=0.41$ for the AGNs with $\Gamma=1.8$ and \textcolor{blue}{Zappacosta et al. (submitted, 2017)} compute the mean $R_f$ equal 0.41 for sample at redshift $0<z<2.1$. These values are within the range of modeled $R_f$ associated with the NSDs: the reflection parameter increases from 0.13 at z = 0 to 0.58 at z = 4.0.
\item The evolution of AGN environment with redshift depends on obscuring medium. As the amount of obscuration increases, the AGN fraction of a given $N_{\text{H}}$ bin in the local universe deviates more with comparison to higher redshift $z=2$.
\item The AGNs with starburst regions alone are not sufficient to produce the observed CXB peak. In addition, 20, 20, and 60 per cent of unobscured, CN, and CK type of high luminosity AGNs (quasars) are required, respectively. The predicted lower limit on CK quasars from the CXB model is 40 per cent.
\item The predicted peak of the SED has a better match with observation as NSDs get more compact. Compact NSDs produce a 1.46 photon index in 2-10 keV band, while the extended NSDs favor higher photon index $\sim 1.52$ \citep{marshall80,deluca04,moretti09,cappelluti17}.
\item With comparison to the observational sample of CDF-S \citep{lehmer12}, the predicted number counts of unobscured AGNs in 2-8 keV band are overestimated. However, the number counts of low obscured AGNs at higher flux ($S>10^{14.7}$ erg s$^{-1}$ cm$^{-2}$) and moderately obscured AGNs are in fair agreement. The estimated total AGN number counts in 8-24 keV band are also consistent with the \textit{NuSTAR} observations \citep{harrison16}.
\item The predicted fraction of obscured ``quasar-like'' AGNs is in reasonable agreement with observations. This supports the possibility that nuclear starburst discs can be an important source of obscuration in Seyfert galaxies at intermediate redshift. \citep{ballantyne08,gohil17}.
\end{itemize}

\section*{Acknowledgement}
This work is supported by NSF award AST 1333360. The authors thank the referee for useful comments.

\newpage
\bibliographystyle{mnras}
\bibliography{refs/ref_a,refs/ref_b,refs/ref_c,refs/ref_d,refs/ref_e,refs/ref_f,refs/ref_g,refs/ref_h,refs/ref_i,refs/ref_j,refs/ref_k,refs/ref_l,refs/ref_m,refs/ref_n,refs/ref_o,refs/ref_p,refs/ref_q,refs/ref_r,refs/ref_s,refs/ref_t,refs/ref_u,refs/ref_v,refs/ref_w,refs/ref_x,refs/ref_y,refs/ref_z,refs/ref_books}{}



\bsp 

\label{lastpage}
\end{document}